\documentclass{ieeeaccess}
\usepackage{cite}
\usepackage{amsmath,amssymb,amsfonts}
\usepackage{algorithmic}
\usepackage{graphicx}
\usepackage{textcomp}
\usepackage{hyperref}
\usepackage{array}
\usepackage{mathrsfs}
\usepackage{balance}
\usepackage{soul,color}
\DeclareUnicodeCharacter{2212}{-}
\newcolumntype{C}[1]{>{\centering\let\newline\\\arraybackslash\hspace{0pt}}m{#1}}
\usepackage{multirow}
\def\BibTeX{{\rm B\kern-.05em{\sc i\kern-.025em b}\kern-.08em
    T\kern-.1667em\lower.7ex\hbox{E}\kern-.125emX}}
    
\begin{document}
\history{Date of Publication February 27, 2021, Date of Current Version February 27, 2021.}
\doi{10.1109/ACCESS.2021.DOI}

\title{CardioXNet: A Novel Lightweight Deep Learning Framework for Cardiovascular Disease Classification Using Heart Sound Recordings}
\author{\uppercase{Samiul~Based~Shuvo}\authorrefmark{1*},
\uppercase{Shams~Nafisa~Ali\authorrefmark{1*}, \uppercase{Soham~Irtiza~Swapnil}\authorrefmark{1}, \uppercase{Mabrook S. Al-Rakhami}\authorrefmark{2} and \uppercase{Abdu Gumaei}\authorrefmark{2, 3}}}

\address[1]{Department of Biomedical Engineering, Bangladesh University of Engineering and Technology (BUET), Dhaka-1000, Bangladesh.}
\address[2]{Research Chair of Pervasive and Mobile Computing, Information Systems Department, College of Computer and Information Sciences, King Saud University, Riyadh 11543, Saudi Arabia.}
\address[3]{Computer Science Department, Faculty of Applied Science, Taiz University, Taiz, Yemen.}

\tfootnote{The authors are grateful to the Deanship of Scientific Research, King Saud University for funding through Vice Deanship of Scientific Research Chairs.}

\markboth
{Shuvo \headeretal: Lightweight DL Framework for Cardiovascular Disease Classification}
{Shuvo \headeretal: Lightweight DL Framework for Cardiovascular Disease Classification}

\corresp{Corresponding author: Mabrook S. Al-Rakhami (e-mail: malrakhami@ksu.edu.sa).}

\begin{abstract}
The alarmingly high mortality rate and increasing global prevalence of cardiovascular diseases (CVDs) signify the crucial need for early detection schemes. Phonocardiogram (PCG) signals have been historically applied in this domain owing to its simplicity and cost-effectiveness. In this paper, we propose CardioXNet, a novel lightweight end-to-end CRNN architecture for automatic detection of five classes of cardiac auscultation namely normal, aortic stenosis, mitral stenosis, mitral regurgitation and mitral valve prolapse using raw PCG signal. The process has been automated by the involvement of two learning phases namely, representation learning and sequence residual learning. Three parallel CNN pathways have been implemented in the representation learning phase to learn the coarse and fine-grained features from the PCG and to explore the salient features from variable receptive fields involving 2D-CNN based squeeze-expansion. Thus, in the representation learning phase, the network extracts efficient time-invariant features and converges with great rapidity. In the sequential residual learning phase, because of the bidirectional-LSTMs and the skip connection, the network can proficiently extract temporal features without performing any feature extraction on the signal. The obtained results demonstrate that the proposed end-to-end architecture yields outstanding performance in all the evaluation metrics compared to the previous state-of-the-art methods with up to 99.60\% accuracy, 99.56\% precision, 99.52\% recall and 99.68\% F1- score on an average while being computationally comparable. This model outperforms any previous works using the same database by a considerable margin. Moreover, the proposed model was tested on PhysioNet/CinC 2016 challenge dataset achieving an accuracy of 86.57\%. Finally the model was evaluated on a merged dataset of Github PCG dataset and PhysioNet dataset achieving excellent accuracy of 88.09\%.
The high accuracy metrics on both primary and secondary dataset combined with a significantly low number of parameters and end-to-end prediction approach makes the proposed network especially suitable for point of care CVD screening in low resource setups using memory constraint mobile devices.
\end{abstract}


\titlepgskip=-15pt

\maketitle

\section{Introduction}
\PARstart{C}{ardiovascular} diseases (CVDs), taking away millions of human lives every year, are inciting major concerns in the global healthcare landscape. In 2016 alone, 17.9 million people, almost 31\% of entire global deaths occurred from CVDs~\cite{CVDCause70:online}. In the same vein, lately an alarmingly upward trend of CVD-orinted deaths has been observed in the developing and under-developed countries~\cite{bowry2015burden}. Therefore, with a view to slackening the current pace of death toll and minimizing the adverse impacts of these diseases, early diagnosis and treatment of CVDs are of paramount importance. 
\par
Heart, one of the most important organs of the human body, produces distinct sounds during its course of mechanical activity. Since the occurrence of a specific type of disease alters the heart functionality in a definite manner, the auscultations also change accordingly and thus, they have been historically utilized for screening CVDs~\cite{bourouhou2019heart}. Although Lub (S1) and Dub (S2) play pivotal role for the detection of cardiac anomalies~\cite{bourouhou2019heart}, sometimes the irregular variants, third heart sound (S3), fourth heart sound (S4) and murmurs are also taken into account while making pathological inference using the stethoscope~\cite{varghees2014novel}. The simple, non-invasive nature of this auscultation based diagnosis strategy have made it the most popular and attractive choice of the cardiologists for performing the initial investigation.~\cite{alaskar2019implementation}. Besides auscultation, multiple other cardiac signals involving a wide range of advanced methods such as, electrocardiogram (ECG), angiography, echocardiography, myocardial perfusion imaging (MPI), cardiac computed tomography (CCT), cardiovascular magnetic resonance (CMR), carotid pulse graph, apex cardiogram etc. are being utilized as modern diagnostic tools for effective screening of CVDs as they vividly reflect the overall transthoracic physiological conditions of the cardiovascular system~\cite{chakir2018phonocardiogram, akram2019quality}. However, in situations where these advanced methods are scarce, lung auscultation stands out as a simple and reliable medium for detecting CVDs. Nevertheless, even for an expert physician, it is quite challenging to readily detect CVDs just by listening to the auscultation. This situation is further exacerbated owing to subjective dissimilar interpretation of the same auscultation by the physicians~\cite{alam2010cardiac}. The disproportionate number of skilled medical professionals with respect to the total population adds up to the situation and further retard the initial diagnostic speed. In this scenario, artificial intelligence-empowered automated cardiac screening systems on the basis of PCG classification can play a pivotal role to assist the physicians in their decision-making process. 

\par
\Figure[t!](topskip=0pt, botskip=0pt, midskip=0pt)[width=\linewidth]{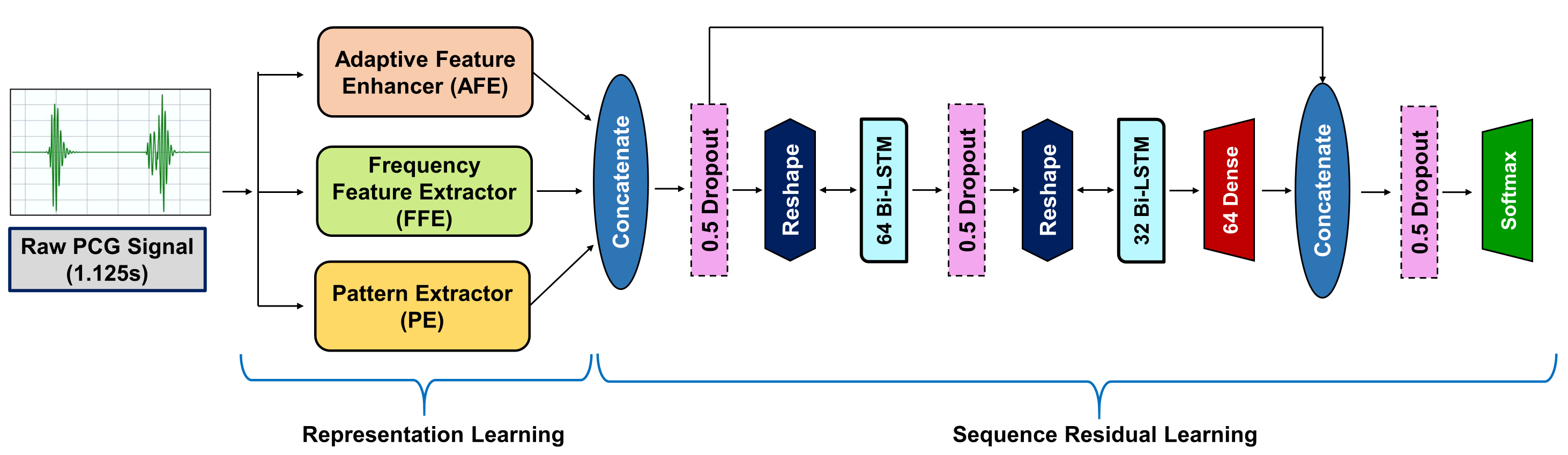}
{An overview architecture of the proposed CardioXNet consisting of 2 phase learning i.e., representation learning and sequence residual learning. The microstructure of the internal modules (AFE, FE and PE) under representation learning phase are shown in figures~\ref{PE},~\ref{AFE}.\label{fig1}}

Since phonocardiogram (PCG), the visualization of heart sound on graphical waveform, can be easily processed to extract essential discriminatory features for the identification of cardiac anomalies~\cite{ismail2018localization}, monitoring heart condition via PCG is becoming an increasingly popular clinical practice. With the amalgamation of machine learning (ML), state-of-the-art networks and advanced audio processing techniques, the unpropitious manual screening can be replaced with automated classification frameworks for prompt large-scale predictions. Therefore, numerous research works have explored the domain of automated PCG classification over the years~\cite{deng2016towards, deperlioglu2020diagnosis, humayun2020towards, zhang2017heart, latif2018phonocardiographic, xiao2020heart, dokur2009feature, patidar2015automatic, khan, oh2020classification,li2020fusion, choi2008detection, uuguz2012biomedical, ghosh2019automated, ghosh2020automated, fahad2018microscopic, m12, m1}. In the literature, a distinct trend of performing 2-step PCG classification can be conspicuously observed. A predominantly trend of performing a segmentation step (localizing lub (S1) and dub(S2) from cardiac auscultation) prior to conducting the classification of the recordings is seen in these works. For materializing the latter step, initially the researchers in the past have utilized statistical analysis and several ML-based algorithms like, Support Vector Machines (SVM)~\cite{bourouhou2019heart, zhang2017heart}, k-Nearest Neighbor (k-NN)~\cite{khan}, Random Forest~\cite{ghosh2019automated}, Hidden Markov Models~\cite{fahad2018microscopic}, Multilayer Perceptron (MLP)~\cite{bourouhou2019heart, dokur2009feature} etc. in conjunction with many signal processing and feature extraction techniques based on time-domain, frequency-domain and time-frequency domain. Among the feature extraction techniques, fast Fourier transform (FFT), short Fourier transform (STFT), discrete wavelet transform (DWT), continuous wavelet transform (CWT), Q wavelet transform (TQWT), chirplet transform (CT), S-transform, Hilbert–Huang transform (HHT), TF decomposition, Empirical mode decomposition (EMD), Mel-Frequency Cepstral Coefficients (MFCCs), Variance Fractal Dimension (VFD) are observed in many works~\cite{ismail2018localization}. Despite being quite promising, most of these ML-based algorithm rely on hand-engineered features in the classifier and thus, are not generalized~\cite{oh2020classification}. Advent of deep learning (DL) architectures namely Convolutional Neural Networks (CNNs), Unsupervised Pre-trained networks (UPNs), Recurrent and Recursive neural networks (RNNs) and their application in the the domain of PCG classification have intuitively resolved both the generalization and the accuracy concern by utilizing the inherent self-learning competency of these networks~\cite{latif2018phonocardiographic, xiao2020heart}. However, these deep networks attain the automated feature extraction capability only after going through a computationally complex extensive training phase with a significantly large dataset~\cite{shuvo2020}. The resource-intense requirement of DL-based frameworks makes them unbefitting to be deployed in the low-resource point-of-care locations of the developing and under-developed countries. Therefore, to implement in mobile platforms and wearable devices, reduction of network parameters along with optimized memory footprint is of paramount importance. A handful of cross-domain studies have investigated the issue and introduced a few advanced strategies such as lightweight networks~\cite{hussain2018swishnet, sandler2018mobilenetv2}, weight quantization~\cite{seo2019efficient} and low precision computation techniques~\cite{hubara2017quantized}. Recently, some studies have successfully implemented these concepts in various biomedical applications including ECG classification~\cite{wu2020ecg}, respiratory disease classification~\cite{shuvo2020} and most importantly, PCG classification~\cite{xiao2020heart}.
\par 
 In this paper, we propose CardioXNet, a novel lightweight CRNN framework with two learning scheme i.e., representation learning and sequence residual learning for automatic detection of CVDs from PCG signal as shown in Figure~\ref{fig1}. Representation learning includes adaptive feature extractor, frequency feature extractor and pattern extractor which combinedly extract time-invariant salient features from the raw audio of the Github PCG database~\cite{khan}. Sequence residual learning contains bidirectional connections and is used for extracting temporal features. The use of these schemes enables the network to classify efficiently and converge rapidly on the raw PCG signals. The major contribution of this work is the automatic end to end classification of valvular heart disease from PCG signals using a lightweight CRNN network with no manual feature extraction or preprocessing steps like segmentation, augmentation or replication.The capability of the proposed deep lightweight CRNN network to extract salient features directly from PCG signals with minimal training parameters and memory is the main feature of this work. Another highlight is the high performance shown in all matrices. The marked improvement of performance is testament to the effectiveness of feature extraction and classification capability of the proposed combination of 1D squeeze excitation and CRNN networks.   
\par
The rest of the paper is organized as follows: In Section \ref{RW}, we have discussed the physiological origin of cardiac ausculation and presented an overview of the previous studies on detection of CVDs using PCG siganl. Deep learning network fundamentals are discussed in Section \ref{basics}; a detailed description of our proposed architecture, along with the intuitions behind designing it is presented in Section \ref{II}; the implementation settings, evaluation criteria and experimental results are discussed with reference to the literature in Section \ref{III};  the future prospects of the work along with the existing limitations are mentioned in Section \ref{IV} and finally, we conclude our study in Section~\ref{V}.

\section{Background} \label{RW}
In the literature, numerous works have been reported on multi-faceted PCG classification. Among them, a vast majority of the existing works have focused on performing normal vs. pathological (abnormal) prediction of the PCG~\cite{bourouhou2019heart, alaskar2019implementation, humayun2020towards, zhang2017heart, latif2018phonocardiographic, xiao2020heart}. Again, some have explored the domain of multi-class (normal, artifact, extrasystole, extrahls, murmurs) cardiac anomaly prediction~\cite{chakir2018phonocardiogram, deng2016towards, deperlioglu2020diagnosis} while some have done multi-class (innocent, systolic, diastolic, pansystolic, ejection) classification of murmurs, the most efficient indicator of CVD incidence~\cite{m12, m1}. Being motivated by the unparalleled success of the aforementioned schemes, several research works from all over the world are being geared towards direct classification of cardiovascular diseases utilizing cardiac auscultation. By elucidating PCG data, to date researchers have successfully classified many CVDs such as, coronary artery disease (CAD), atrial fibrillation (AF), aortic stenosis (AS), dilated cardiomyopathy (DC), mitral stenosis (MS), papillary muscle dysfunction (PMD), mitral regurgitation (MR), atrial septal defect (ASD), ventricular septal defect (VSD), mitral valve prolapse (MVP) etc.~\cite{dokur2009feature, patidar2015automatic, khan, oh2020classification,li2020fusion, choi2008detection, uuguz2012biomedical, ghosh2019automated, ghosh2020automated}.

\subsection{Physiological Origin of Cardiac Auscultation}
Cardiac auscultation is a complex sound signal generated by the mechanical activities of the heart. It contains the impact of vibration of the valvular opening and closing, systolic and diastolic movement of muscles and tendons and the internal blood perfusion~\cite{dwivedi2018algorithms}. However, there are two types of fundamental heart sounds (FHSs). At the onset of the systolic phase, when the atrioventricular (AV) mitral and tricuspid valves close, the first FHS, “lub” (S1) can be heard~\cite{mason2000listening}. The definitive abnormality in S1 might indicate the occurrence of MS, AF and ASD~\cite{HS}. S1 is followed by the second FHS, “dub” (S2) at the beginning of the diastolic phase from the closure of the pulmonic and aortic valves. This relatively high-pitched sound demonstrates higher frequency~\cite{barma2014quantitative} in case hypertension and AS during an event of cardiac abnormality. 
\par 
In normal cardiac condition, distinct pattern of systolic and diastolic period can be observed. Nevertheless, 3rd and 4th heart sound, respectively S3 and S4, murmurs, gallops, clicks might occur because of CVD incidence~\cite{dwivedi2018algorithms}. While S3 mostly indicates reduced systolic function and congestive heart failure (except for pregnant ladies, athletes and young people); S4 is an clear indicator of diastolic dysfunction. The audible whooshing sounds i.e., murmurs might be generated because of the stiffening of the heart valves, narrowing of the blood vessels and turbulent flow~\cite{barma2014quantitative}. Murmurs are very significant for classifying CVDs, as their diverse pitch and frequency pattern represent many CVDs including MR, AR, AS, MS etc.~\cite{HS}. However, innocent murmurs are sometimes produced depending on the physiological condition of the individual.

\subsection{Literature Review}
As mentioned earlier, the automatic detection of CVDs and heart valve disorders (HVDs) have been explored in multiple previous works. Some of these works have utilized publicly available data while some implemented the task with in-house datasets. In this subsection, we will provide a detailed overview of some of the prominent works in this domain. 
\par
Employing divergence analysis on 50 wavelet features with MLP classifier, Dokur \textit{et. al}~\cite{dokur2009feature} has achieved up to 99\% accuracy in determining discriminatory features of VSD, MR, AR, AS, MS, diastolic rumble, summation gallop, aortic insufficiency and some other cardiac abnormalities using a self-constructed multi-source data. In \cite{patidar2015automatic}, septal defects were detected along with other heart defects like valvular defects, ventricular hypertrophy, constrictive pericarditis etc from publicly a available PCG database using TQWT based sub-bands decomposed from segmented heart disease.The classification was performed using least squares support vector machine (LS-SVM) algorithm and achieved overall accuracy of 99.03\%. Using the Github PCG database~\cite{khan}, a multivariate comprehensive analysis for detecting 5 classes, i.e., MS, AS, MR, MVP, N has been carried out in~\cite{khan}. Discrete Wavelets Transform (DWT) and Mel Frequency Cepstral Coefficient (MFCCs) have been employed as feature extraction technique while SVM, deep neural network (DNN) and centroid displacement based K-NN classifier have been adopted as classifier. The features and classifiers are attempted both single-handily and in an ensemble manner through rigorous experimentation. Merging the DWT and MFCC features, centroid displacement based K-NN, DNN and SVM respectively yields an accuracy of 97.4\%, 92.1\% and 97.9\%~\cite{khan}. In the same vein, using the same dataset, a recent study has proposed WaveNet, a novel 1D CNN network having residual blocks with dilated 1D CNN and multiple skip connections~\cite{oh2020classification}. Upon 10-fold cross-validation, they have achieved training accuracy of 97\% and approximately 90\% validation accuracy. Along with sensitivity 92.5\% and specificity 98.1\%, the model exhibits satisfactory performance. A comprehensive basis for CAD diagnosis has been provided in~\cite{li2020fusion} employing a novel multi-domain feature fusion framework. For the in-house dataset used, the concerted feeding of the selected MFCC features with the deep features into a MLP classifier has obtained a promising accuracy of 90.43\%. 
\par
\begin{figure*}[t]
  \includegraphics[width=\textwidth,keepaspectratio=true]{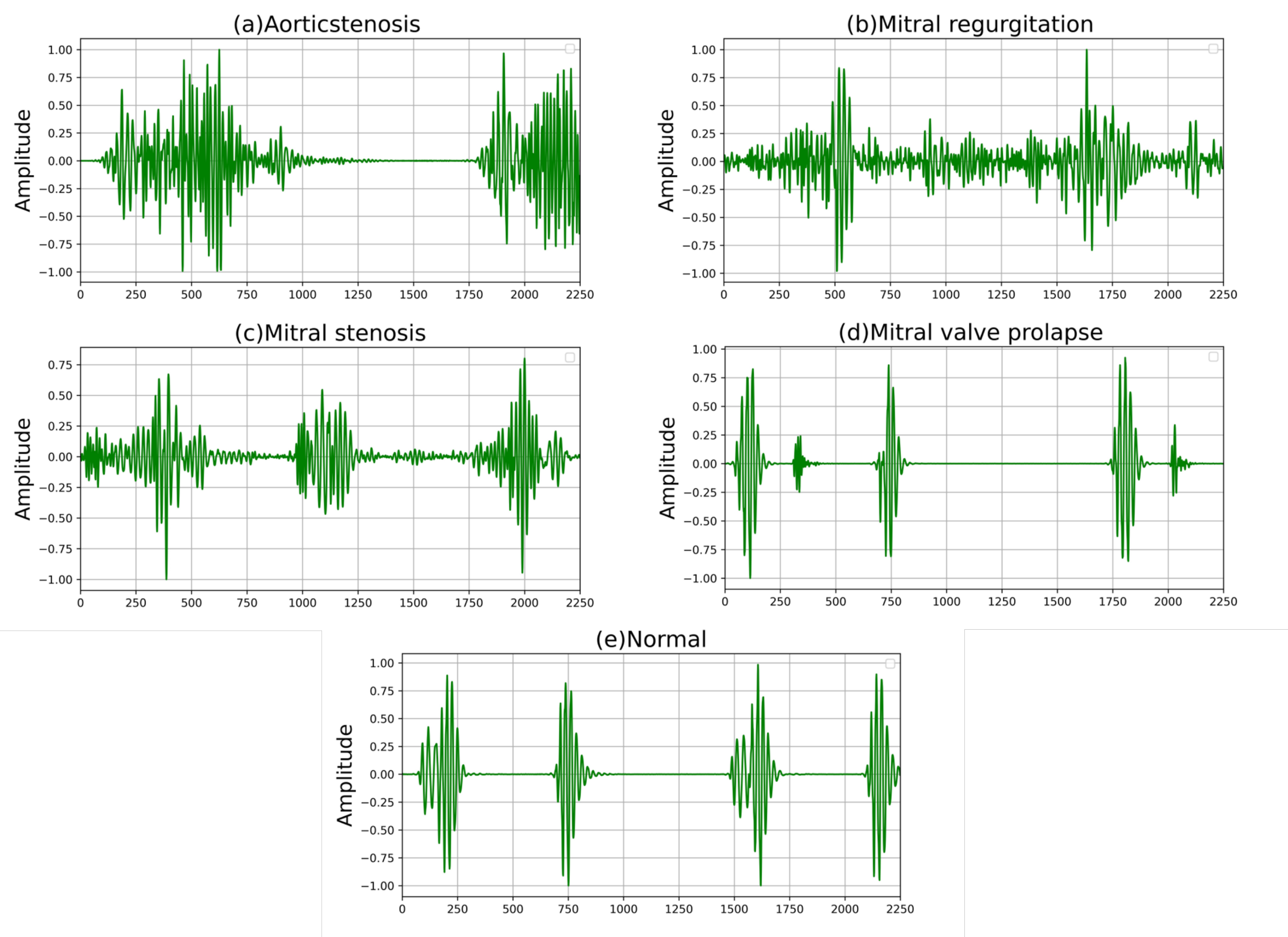}
  \caption{Waveform of the existing CVD classes on PCG dataset. (a) Aortic stenosis (AS), (b) Mitral regurgitation (MR), (c) Mitral stenosis (MS), (d) Mitral valve prolapse (MVP) and (e) Normal.}
  \label{wav}
\end{figure*}

Several studies have utilized PCG signals for classifying heart valve disorders (HVDs). In \cite{choi2008detection}, an in-house dataset of 120 PCG signals was used for HVD classification using wavelet packet(WP) decomposition and SVM. The classification task was performed on 2 classes (Normal and HVDs) achieving specificity of 96.67\%. Another work~\cite{uuguz2012biomedical} involving in-house dataset, has classified  normal, pulmonary and mitral stenosis heart valve diseases utilized discrete fourier transform (DFT) and Burg autoregressive (AR) spectrum analysis for feature extraction. Using Principal Component Analysis (PCA) for dimensionality reduction and ANN for classification, the work has achieved maximum accuracy score of 95\%. Maintaining the legacy, Ghosh\textit{et. al}~\cite{ghosh2019automated} has classified the HVD classes i.e., MR, AR, AS, N using magnitude and phase features, wavelet synchrosqueezing transform (WSST) based time-frequency matrix with a Random Forest classifier on the Github PCG database~\cite{khan} dropping the MVP class. The proposed paradigm has demonstrated the mean accuracy to be 95.13\%. In a later work by the same author~\cite{ghosh2020automated}, using Chirplet transform (CT), local energy (LEN) and local entropy (LENT) features in conjunction with a composite classifier, an overall accuracy of 98.33\% was obtained on the same dataset.
\par In all of these works, separate preprocessing steps have been performed for feature extraction. However, in one of the recent studies, a CNN-LSTM model ~\cite{alkhodari2021convolutional} has been used for direct classification using the PCG dataset with 1D wavelet smoothing, achieving 99.32\% accuracy. The use of such models without manual preprocessing steps are vital for any end to end mobile health care devices.

\section{Materials}
\subsection{Dataset Description}
In this study, PCG recordings used in the article~\cite{khan} has been primarily utilized .The recordings were collected from various sources like books and websites and contained a total number of 1000 PCG recordings in .wav format in five different classes i.e., Normal (N), Aortic stenosis (AS), Mitral regurgitation (MR), Mitral stenosis (MS), Mitral valve prolapse (MVP). Each of the classes has 200 recordings for roughly 3s. All the recordings are sampled at 8 kHz. Since the lowest signal length present in this dataset is 1.125s, all the recordings are truncated from the start of the recording up to 1.125s. Figure ~\ref{wav} illustrates the waveform for each of the disease classes.
\par
Additionally, to validate the PCG classification performance of our proposed network, the openly accesible ``PhysioNet/CinC Challenge 2016 dataset~\cite{data_physionet}'' is incorporated as a secondary dataset that contains a total number of 3240 PCG recordings within six separately labeled datasets. Seven different research groups have concertedly collected the PCG signals, at either clinical or nonclinical settings. The recordings are originally sampled at 2 kHz and have varying duration (5s-120s). Nevertheless, this dataset does not contain disease-based annotation and contains only 2 classes (normal, abnormal), where the normal annotated 2575 recordings refer to healthy subjects with no valvular defects and the remaining 665 abnormal recordings indicate pathological cases like arrhythmia, coronary heart disease, valvular stenosis, mitral regurgitation etc. For maintaining analogy with the primary dataset and preserving fidelity of the proposed model, only the first 1.125s from each of the PCG recordings of PhysioNet dataset were taken as well.

\subsection{Preprocessing}
\begin{figure*}[h]
  \includegraphics[width=\linewidth]{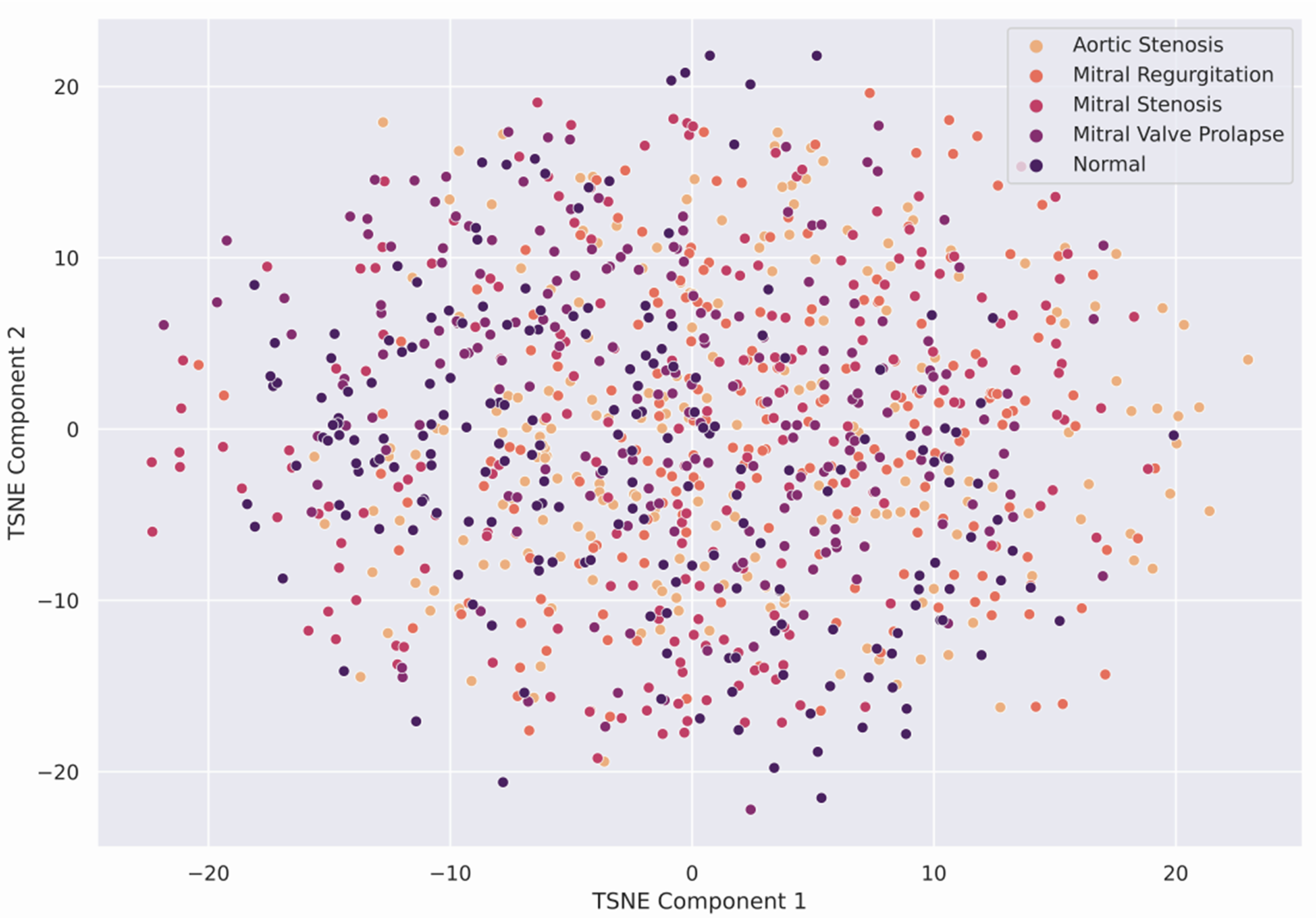}
  \caption{t-Stochastic Neighbour Embedding (t-SNE) visualization  of normal class and four CVDs class Aortic stenosis (AS), Mitral regurgitation (MR), Mitral stenosis (MS) Mitral valve prolapse (MVP) heart sound data.}
  \label{TSNE}
\end{figure*}

All the PCG signals are resampled at 2 kHz that ensures the preservation of important heat sound frequency components as well as  lowers the computational cost. The signals are also amplitude normalized to ensure consistency and homogeneity by nullifying the effect of acquisition device/sensor variation. 
The visible clusters in Figure ~\ref{TSNE} shows feature distributions corresponding to different CVD class along with normal class .

\section{Network Fundamentals}  \label{basics}
\subsection{Convulational Neural Network (CNN)}
Convolutional neural networks, a variant of state-of-the-art artificial neural networks, mainly combine three architectural generalization ideas by ensuring different invariance techniques such as, shared weights, local receptive fields and the subsequent presence of sub-sampling of spatial or temporal channels with a view to reducing the complexity of the network model. Each of the CNN layer consists of a number of neural units and receives a set of extracted features as input from the previous layer. CNNs for images, speech, and time-series signals have different sets of kernel formation and they map the input into different types of features from each location~\cite{lecun1995convolutional}.
\subsubsection{1D CNN}
In spite of being analogous to the regular neural network (ANN), 1D CNN generally takes raw sequential data i.e., time-series data like PCG, EEG as input. The main advantage of 1D CNN over ANN is that it can show excellent performance on time-series data without engaging any time or resource intensive feature extraction steps.
\par
Assuming the input of 1D CNN is a tensor representing the audio waveform denoted as X,  feature extraction of 1D convolution can be defined as,
\begin{equation}
T = F(X | \Theta) = f_L(...f2(f1(X | \Theta_1) | \Theta_2) | \Theta_L)
\end{equation}
Here, L is the number of hidden layers, $\Theta$ is a set of parameters that maps the input to the prediction T.
\par
The l-th convolutional layer's operation can be defined as:
\begin{equation}
   T_l = f_l(X_l | \Theta_l) = h(W \otimes X_l + b), \Theta_l = [W, b] 
\end{equation}
Here, $\otimes$ denotes the convolution operation,$ X_l $ is a two-dimensional input matrix of
N feature maps, W is a set of N one dimensional kernels (receptive fields) used for
extracting a new set of features from the input array, b is the bias vector, and h (·) is the
activation function. The shapes of $ X_l $, W and $ T_l $ are respectively (N, d), (N, m) and (N, d$-$m + 1)~\cite{abdoli2019end}.
\subsubsection{2D CNN}
2D CNN is generally used for extracting spatial features using its kernel. It has multiple hidden layers, each layer having a number of neural units. The neural unit is basically a function connecting inputs and outputs. The input function can be denoted as ~\cite{chen2020classification}:
\begin{equation}
    x_k^n=\sum_{n=1}^{N_{n-1}} conv2D(w_{ik}^n,s_i^{n-1})+b_k^n
\end{equation}
here,$s_i^{n-1}$represents
the output of the ith neuron of (n-1)th layer and $b_k^n$
and $(w_{ik}^n$
are the bias and kernel of
the kth neuron in the nth layer,accordingly.conv2D represents a two-dimensional regular convolution with valid padding. The input function of the kth neuron can be denoted as~\cite{chen2020classification}:
\begin{equation}
    y_k^n=f\Bigg(\sum_{n=1}^{N_{n-1}} conv2D(w_{ik}^n,s_i^{n-1})+b_k^n\Bigg)
\end{equation}
\subsection{Recurrent Neural Network (RNN)}
RNN is a special kind of ANN model where the input samples contain more inter-dependencies. RNN has a significant specialisation at storing the information of the past time steps. In other words, the parameters available at time $t_1$+1 will be affected by output produced at time $t_1$~\cite{supratak2017deepsleepnet}. RNN is mainly used for sequential and dynamic information processing and thus, it is very useful for multiple types of audio signal processing i.e., music and speech processing, pathological auscultation processing etc. The Generalized delta rule~\cite{huk2012backpropagation} and energy minimization function are two of the major algorithms used for RNN training~\cite{kalchbrenner2013recurrent}.
\subsubsection{Long-Short Term Memory}
\begin{figure}[t]
\centering
  \includegraphics[width=\columnwidth]{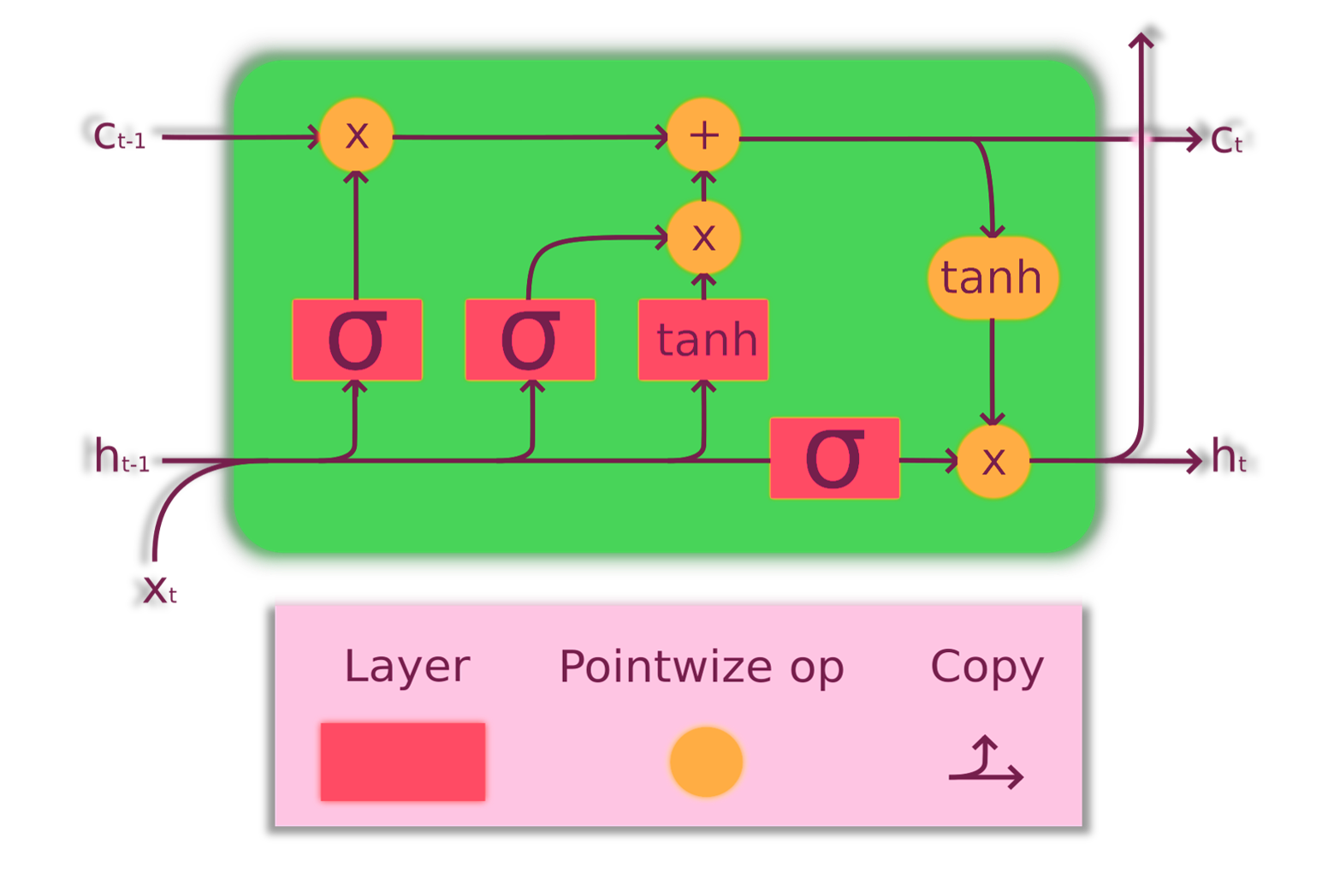}
  \caption{The structure of the LSTM cell.}
  \label{LSTM cell}
\end{figure}

The Long-Short Term Memory (LSTM), proposed by Hochreiter and Schmidhuber in 1997 is a primary building block of RNN. It has been substantially explored for sequence learning. A standard LSTM cell (see Figure~\ref{LSTM cell}) consists of three distinct gates and three main cell-memory networks. The respective equations for each of these gates and networks are presented below:

\par
 Input gate:
 \begin{equation}
  \\i_t =  Sigmoid (W_{xi}x_t +W_{hi}h_{{t−1}} + W_{ci}c_{{t-1}} + bi)\
 \end{equation}
  Forget gate:
 \begin{equation}
 \\f_t =  Sigmoid (W_{xf}x_t + W_{hf}h_{{t−1}} + W_{cf}c_{{t-1}} + bf)\
 \end{equation}
  Output gate:
  \begin{equation}
 \\o_t = Sigmoid (W_{xo}x_t + W_{ho}h_{{t−1}} + W_{co}c_{{t-1}} + bo)\
 \end{equation}
Input cell: 
 \begin{equation}
 \\g(t)\ = tanh(W_{xg}x_t +W_{hg}h_{{t−1}} + b_g)\
 \end{equation}
Memory cell:
\begin{equation}
\\c(t)\ = f_t \odot c_{{t−1 }}+ i_t \odot g_t\
\end{equation}
Output cell:
\begin{equation} 
\\h(t)\ = o_t \odot tanh(c_t)\
\end{equation}

Here, sigmoid and tanh are the activation functions which map the non-linearity of the features~\cite{hochreiter1997long}.
\subsubsection{Bidirectional LSTM}
The main disadvantage of one-directional LSTM is that it is only encompasses the past information. To overcome this limitation of the one-directional LSTM, a bi-directional LSTM  (Bi-LSTM) approach has been introduced by Schuster and Paliwal~\cite{schuster1997bidirectional} for efficiently investigating both the past and the future context. The Bi-LSTM is defined as,
\begin{equation}
\overrightarrow{h} = \sigma(x_tU+\overrightarrow{h}_{{t-1}}W + b_t) 
\end{equation}

\begin{equation}
\overleftarrow{h} = \sigma(x_tU+\overleftarrow{h}_{{t-1}}W + b_t) 
\end{equation}
Here, $\overrightarrow{h}$ is the forward hidden state and $\overleftarrow{h}$ is the backward hidden state. Concatenation of both of these states formed the hidden state at time t .

\subsection{Micro-architecture of Fire module}
The Fire module, composed of two layers i.e., Squeeze and Expansion layer, is a dedicated parameter reduction module introduced in the SqueezeNet architecture~\cite{iandola2016squeezenet}. In the Squeeze layer, 3 × 3 kernels have been replaced with 1 × 1 kernels which results in 9x reduction in the number of parameters. Then, the squeezed features are fed into an expansion layer. The expansion layer consists of a layer with 1 x 1 filters and a layer with 3 x 3 filters and the extracted features of these layers concatenated together in the channel dimension. By maintaining 8x fewer output channels in the squeeze layer compared to the expansion layer, squeeze ratio (SR) of 0.125 has been achieved. 

\section{Network Architecture}  \label{II}
The proposed network namely CardioXNet is constructed in a modular way, employing two major learning paradigms i.e., representation learning and sequence residual learning. This architecture is designed for effectively classifying short unsegmented PCG recordings. Figure ~\ref{fig1} portrays the high-level network architecture.
\par
\begin{figure}[h!]
  \includegraphics[width=\linewidth]{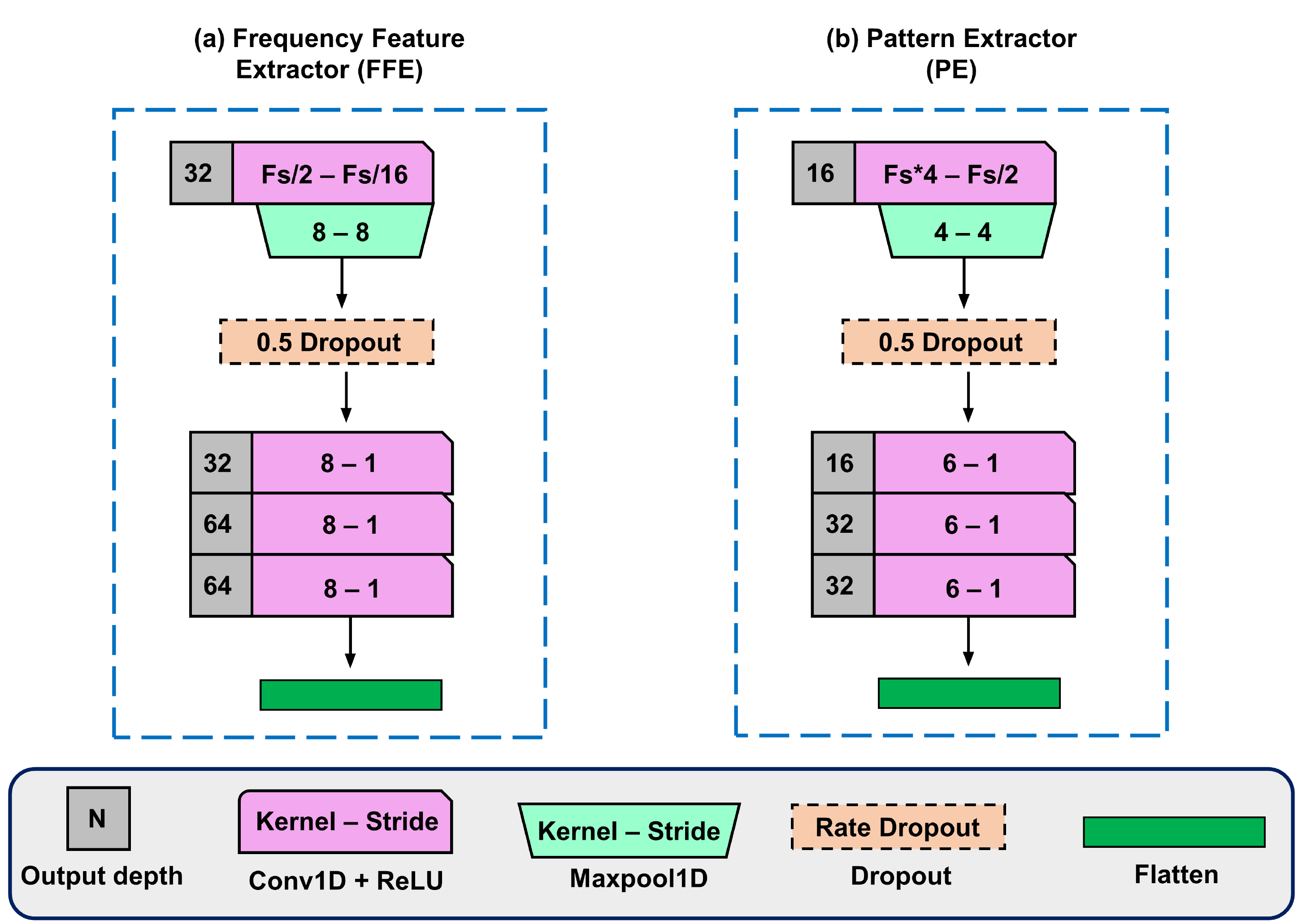}
  \caption{Micro structure of (a) Frequency Feature Extractor (FFE); (b) Pattern Extractor (PE). Fs represents sampling frequency. }
  \label{PE}
\end{figure}
\par
\begin{figure}[h!]
  \includegraphics[width=\linewidth]{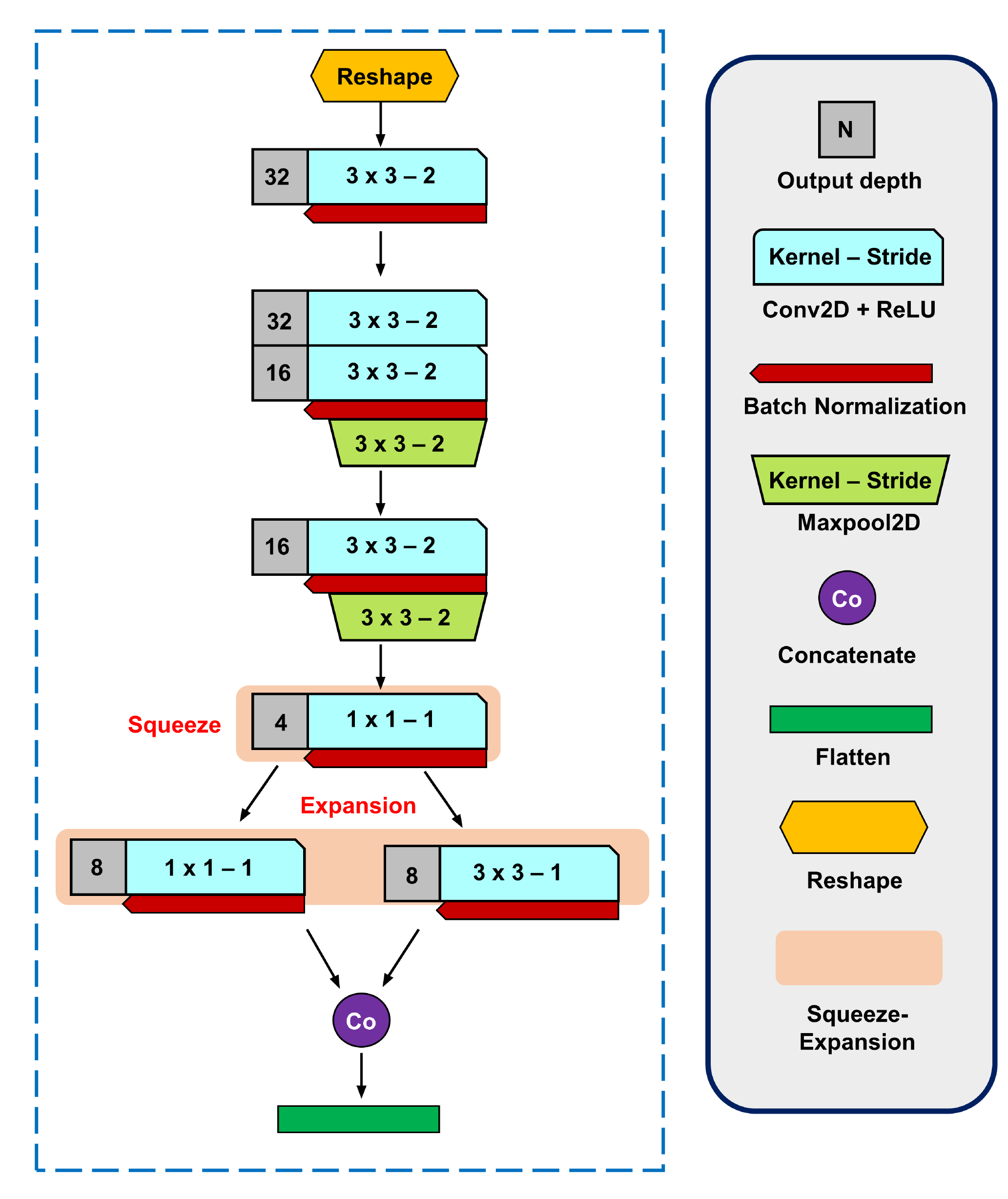}
  \caption{Micro structure of Adaptive Feature Enhancer (AFE).}
  \label{AFE}
\end{figure}
\par 

\subsection{Representation learning}
This part is trained with a view to extracting time-invariant features from each of the raw PCG signals. 
It consists of three parallel CNN pathways namely, Frequency Feature Extractor (FFE), Pattern Extractor (PE) and Adaptive Feature Enhancer (AFE)~\cite{supratak2017deepsleepnet,iandola2016squeezenet}. 
\subsubsection{Frequency Feature Extractor (FFE)}
FFE consists of four 1D convolutional layers and two max-pooling layers with the primary filter size of sampling frequency (Fs) × 4 and stride size set to Fs/2 for the 1D convolutional (conv1) layer to capture the frequency components. The subsequent filter and stride sizes are chosen by rigorous hyperparameter tuning (see Figure~\ref{PE}).
\subsubsection{Pattern Extractor (PE)}
Similar to FFE, PE also consists of four 1D convolutional layers and two max-pooling layers. However, fine-grained convolution with filter and stride size set to Fs/2, and Fs/16 respectively are used which helps to recognise the appearance of various PCG pattern. With the help of hyperparameter tuning, the filter and stride sizes are chosen (see Figure~\ref{PE}). 

\subsubsection{Adaptive Feature Enhancer (AFE)}
The input sequences are reshaped into a 2D tensor and fed to AFE. AFE consists of 2D convolutional layers, batch-normalization, max-pooling layers and squeeze-expansion layers, inspired by the Fire module of SqueezeNet architecture~\cite{iandola2016squeezenet}. Batch normalization layers helps to stabilize and speed up the training process~\cite{ioffe2015batch} while abstract feature maps generated from the initial convolutional layers are then passed to the squeeze-expansion layers which results in lowering of parameter count by manifolds(see Figure~\ref{AFE}). Apart from efficient feature extraction, it also ensures faster convergence in the training phase, a highly appreciated task from the perspective of training deep networks in mobile platforms.
\begin{figure*}[h!]
  \includegraphics[width=\linewidth]{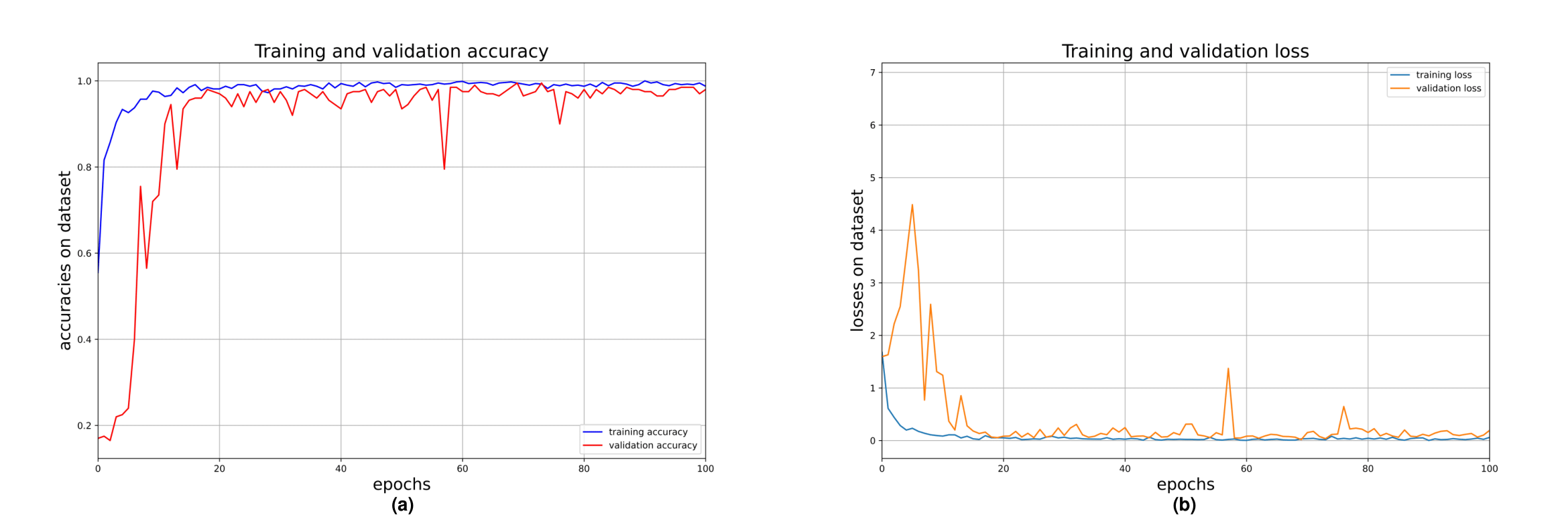}
  \caption{(a) Epoch vs accuracy of the CRNN during training and validation (b) Epoch vs loss of the CRNN during training and validation for 10th cross validation step}
  \label{converge}
\end{figure*}
\begin{figure}[h!]
  \includegraphics[width=\columnwidth]{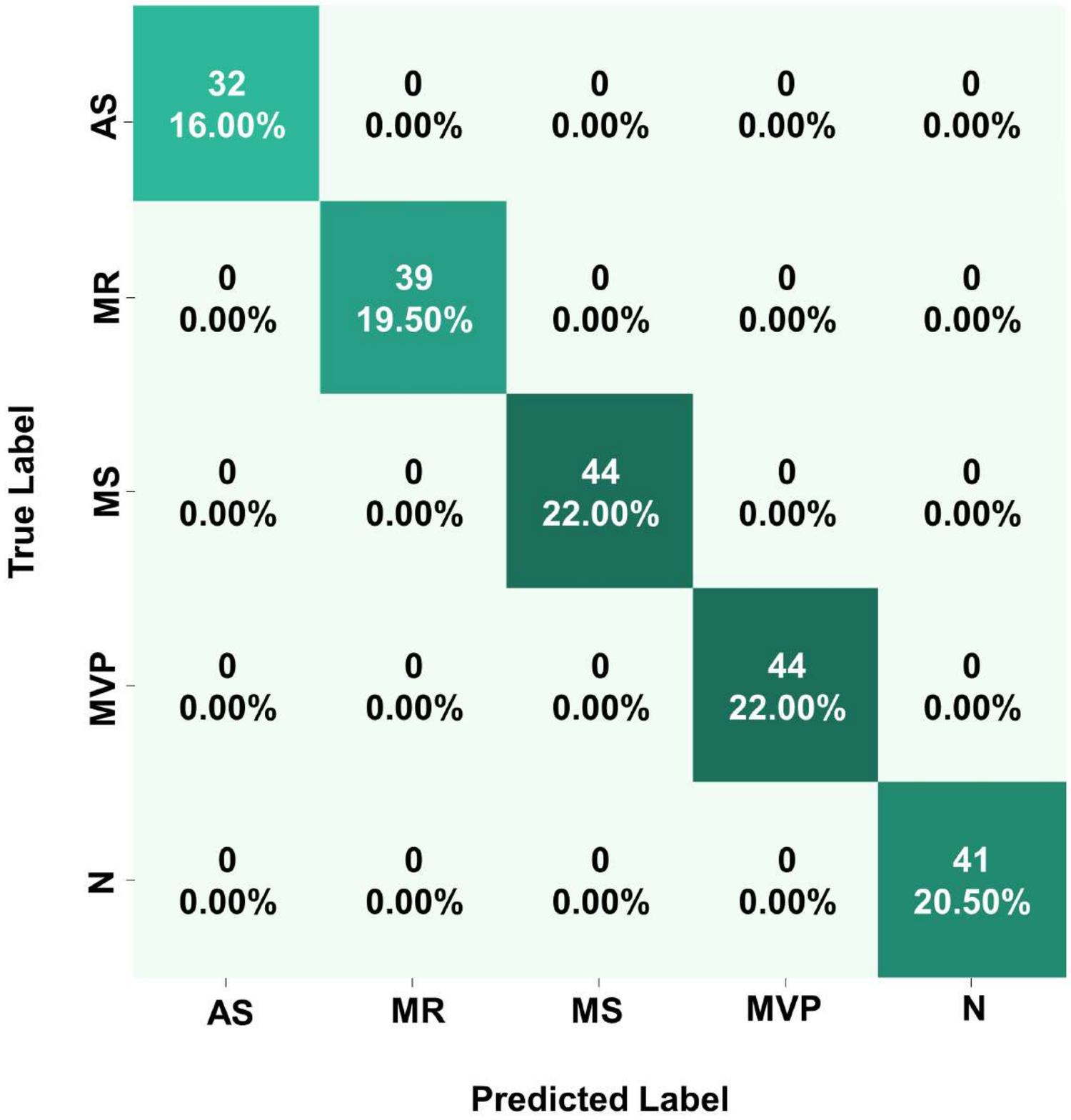}
  \caption{Confusion matrix of the GitHub PCG dataset.}
  \label{Confusion Matrix 1}
\end{figure}
\begin{figure}[h!]
  \includegraphics[width=\columnwidth]{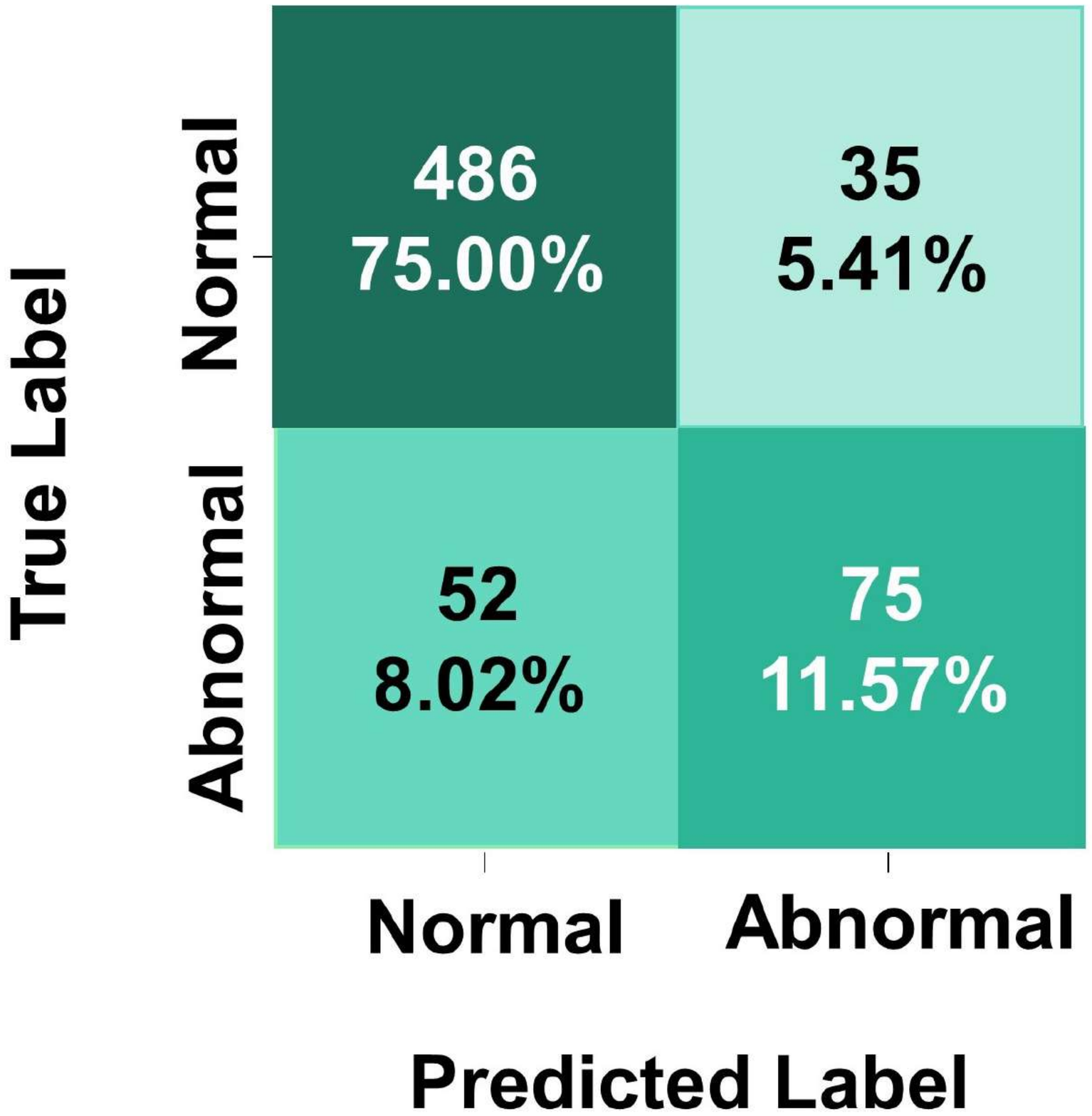}
  \caption{Confusion matrix of PhysioNet/CinC 2016 dataset.}
  \label{Confusion Matrix 2}
\end{figure}
\begin{figure}[h!]
  \includegraphics[width=\columnwidth]{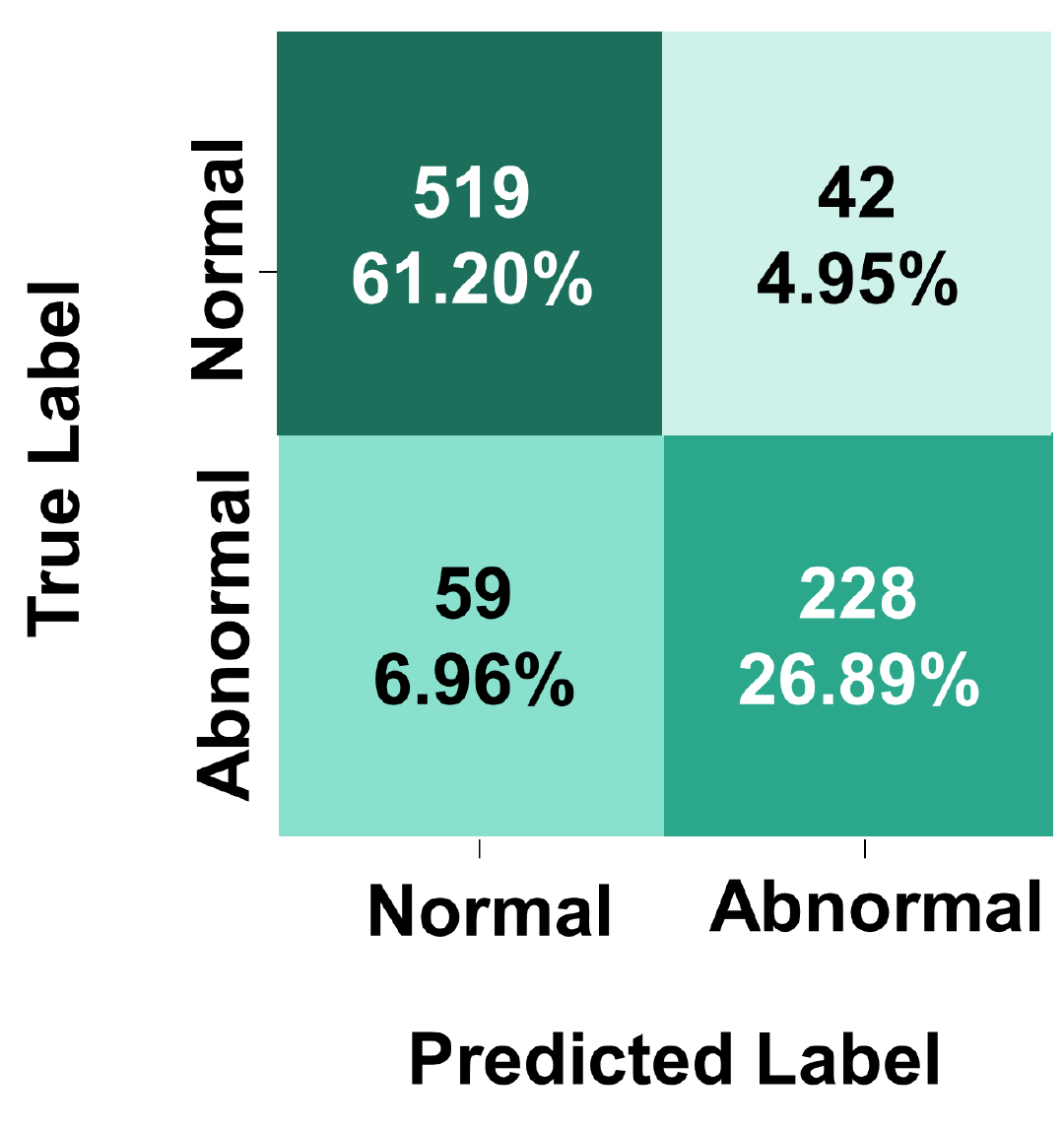}
  \caption{Confusion matrix of the best model on PhysioNet/CinC 2016 and GirHub PCG mixed dataset.}
  \label{Confusion Matrix 3}
\end{figure}
\par Outputs from three CNN paths are concatenated together and forwarded to the sequence residual learning part.
 \begin{equation}
  X\textsubscript{concat} = X\textsubscript{AFE}\oplus X\textsubscript{FFE} \oplus X\textsubscript{PE}
\end{equation}
Here, X\textsubscript{concat} is the concatenated feature vector and X\textsubscript{AFE}, X\textsubscript{FFE},X\textsubscript{PE} are the extracted feature vectors from three CNN paths.

\subsection{Sequence residual learning}
Sequence residual learning is trained to extract the temporal information from the sequence of extracted features in the representation learning part.
Two layers of bi-LSTMs have been employed to learn temporal information which enables the encoding of both past and future information by processing with two independent LSTMs. Since PCG is a time-series data, the application of Bi-LSTM is best suited for capturing the sequential information embedded in them. 
\par
A skip connection has been employed to implement residual functionality and enabling the addition of temporal information and previously extracted features from the CNNs.The concatenated feature vector is fed into a prediction layer with probability nodes, calculated by the softmax function.
\par In all the convolutional layers and LSTM layers, Rectified Linear Unit (ReLU)~\cite{xu2015empirical} activation function has been used, which is defined as:
\begin{equation}
 f(x) = max (0, x), 
\end{equation}

\par It has been employed to diminish the vanishing gradient issue and also contributes to the faster convergence. For minimizing the over-fitting phenomenon, dropout layers have been employed.

\section{Experiments}  \label{III}

\subsection{Evaluation Metrics}
The dataset was divided into training, validation and testing parts with 70\% of the PCG signal used for training the model and 10\% of the signal used for validation and remaining 20\% data used for testing. 10-fold cross validation was performed randomly on the dataset for generailization of the obtained results. Well-known and important performance evaluation metrics such as, Accuracy, Precision , Recall and F1- score are chosen to evaluate the proposed framework quantitatively.

\begin{equation}
~\\Acc=\frac{TP+TN}{TP+FP+TN+FN}\
\end{equation}

\begin{equation}
\\Precision=\frac{TP}{TP+FP}\ 
\end{equation}

\begin{equation}
\\Recall=\frac{TP}{TP+FN}\ 
\end{equation} 

\begin{equation}
\\F1\;score=\frac{2*TP}{2*TP+FP+FN}\ 
\end{equation} 
\subsection{Experimental Setup}
The CardioXNet is designed using Keras and TensorFlow backend and training work has been performed using NVidia K80 GPUs provided by Kaggle notebooks. 
\par
Sparse categorical crossentropy loss and adaptive learning rate optimizer (Adam) with the learning rate of 0.00001 have been engaged in the training process and a batch size of 16 is selected for training and validation purpose.
\begingroup
\setlength{\tabcolsep}{25pt} 
\renewcommand{\arraystretch}{1.5} 
\begin{table*}[h!]
\centering
\caption{Results of 10-fold cross-validation}
 \label{Classification Report}
\centering
\begin{tabular}{|c|c|c|c|c|}
\hline
\textbf{Fold iteration}&	\textbf{Accuracy}&\textbf{Recall}&	\textbf{Precision}&		\textbf{F1-score}\\
\hline
$1$	& $0.995$	&$0.995$&	$0.997$&	$0.996$\\
\hline
$2$	& $0.99$	&$1.00$&	$1.00$&	$1.00$\\
\hline
$3$	& $1.00$	&$1.00$&	$1.00$&$1.00$\\
\hline
$4$	& $0.995$	&$0.996$&	$0.995$ &	$0.995$\\
\hline
$5$	& $0.990$	&$0.984$&	$0.984$&	$0.991$\\
\hline
$6$	& $0.995$	&$0.997$&	$0.995$&	$1.00$\\
\hline
$7$	& $0.995$	&$0.996$&	$0.995$&	$0.995$\\
\hline
$8$	& $1.00$	&$1.00$&	$1.00$&	$1.00$\\
\hline
$9$	& $0.995$	&$0.994$&	$0.994$&	$0.995$\\
\hline$10$	& $0.995$	&$0.996$&	$0.996$&$0.996$\\
\hline
$Mean$& $0.996$&$0.9952$&	$0.9956$&	$0.9968$\\
	\hline
\end{tabular}
\end{table*} 
\endgroup

\begingroup
\setlength{\tabcolsep}{6pt} 
\renewcommand{\arraystretch}{1.5} 
\begin{table}[h!]
\centering
\caption{Classification Report for different classes using CardioXNet}
 \label{Classification Report2}
\centering
\begin{tabular}{|C{1cm}|c|c|c|c|}
\hline
\textbf{Class}&	\textbf{Precision}&	\textbf{Recall}&	\textbf{F1-score}\\
\hline
$AS$	& $0.9733(\pm0.028) $&	$0.9951(\pm0.010)$&	$0.9840(\pm0.014)$\\
\hline
$MR$	& $0.9730(\pm0.023) $	&$0.9807(\pm0.024)$&	$0.9776(\pm0.023)$\\
\hline
$MS$	& $0.9899(\pm0.013)$	&$0.9846(\pm0.021) $&	$0.9842(\pm0.016)$\\
\hline
$MVP$	& $ 0.9949(\pm0.011)$& $0.9696(\pm0.019) $&	$0.9821(\pm0.014)$ \\
\hline
$N$	& $1.0000(\pm0.000)$&$1.0000(\pm0.000)$&	$1.0000(\pm0.000)$\\
\hline
\end{tabular}
\end{table} 
\endgroup
\subsection{Performance of the Proposed Framework}
In this work, the proposed lightweight model has been evaluated on the Github PCG datast, PhysioNet/CinC challenge dataset and on both of these dataset combined. In the following sections a detailed explanation of these results are discussed.
\subsubsection{Performance on GitHub PCG dataset}
The PCG dataset has been classified into 5 different classes using CardioXNet. The proposed model achieved near perfect validation accuracy on the given dataset,  on 10 fold cross validation, showing very high precision and recall scores for all the classes (see Table.~\ref{Classification Report} and Table.~\ref{Classification Report2}). Furthermore, Figure~\ref{converge} (a) depicts that the model converged with great rapidity reaching 100\% training accuracy, while validation accuracy reached 99.5\% within 100 epochs. Additionally, Figure~\ref{converge} (b) provides an insight into the minimization of the categorical cross entropy loss. The incorporation of adaptive feature extractor block with the squeeze and expansion layers mainly attributed to this faster convergence of the model.
\subsubsection{Performance on PhysioNet/CinC 2016 dataset}
To demonstrate the effectiveness of the proposed model, classification between healthy and unhealthy classes were performed on a secondary dataset, namely PhysioNet/CinC 2016 challenge dataset. The confusion matrix of the best performing models for both GitHub PCG dataset and secondary dataset is shown in Figure~\ref{Confusion Matrix 1} and~\ref{Confusion Matrix 2}. From Figure~\ref{Confusion Matrix 2}, we can see, the proposed framework performs quite well on PhysioNet dataset showing around 86.57\%($\pm$2.31) accuracy,93.28\%($\pm$2.10) precision,90.33\%($\pm$1.90) recall and 91.78\%($\pm$1.99) F1 Score. This remarkable performance on a secondary dataset proves the excellent capability of the proposed model to differentiate features without any pre-processing steps like segmentation and augmentation. 
\par The difference in accuracy between the GitHub PCG dataset and PhysioNet dataset might be attributed to the presence of noise and variation in recording devices and variable sources within the PhysioNet dataset. 

\subsubsection{Performance on PhysioNet-GitHub merged dataset}
The generalization potential of the proposed CardioXNet has been tested using a merged dataset of PhysioNet and Github PCG recordings for training, validation and testing. The four diseased class data from GitHub PCG database was merged with unhealthy data from PhysioNet dataset while Normal class data from GitHub PCG dataset was matched with healthy dataset from PhysioNet dataset. Overall the mixed dataset consists of  4240 (3240 + 1000) PCG recordings in which 2757 (2575+200) recordings are normal and 1465 (665+800) are abnormal. The total recordings 
available for training are 2968 while 424 and 848 recordings were respectively, used for testing purpose.Two class classification performance on the mixed dataset is shown in Figure~\ref{Confusion Matrix 3}. CardioXNet showed overall accuracy of 88.09\%, 88.08\% precision, 87.98\% recall and 88.03\% F1 Score. This excellent performance shows the superior feature extraction and classification performance of our proposed model. 
\begingroup
\setlength{\tabcolsep}{26pt} 
\renewcommand{\arraystretch}{1.5} 
\begin{table*}[h!]
\caption{Comparison between previous works and CardioXNet on CVD classification}
\label{Accuracy table}
\begin{center}
\begin{tabular}{|C{3cm}|c|c|c|}
\hline
\textbf{Feature Extraction	}&\textbf{Model}		&	\textbf{Accuracy}	&	\textbf{F1-score}	\\
\hline
	&KNN	&$97.4$	&	$99.2$		\\
\cline{2-4}
$MCC+DWT$~\cite{khan}&DNN		&$97.9$	&	$99.7$	\\
\cline{2-4}
&SVM	&$92.1$	&	$98.3$	\\
\hline
-& WaveNet~\cite{oh2020classification}		&$90.0$	&	-\\
\hline
$WSST$~\cite{ghosh2019automated} & Random Forest & $95.13$ & - \\
\hline
$Chirplet~
Transform$~\cite{ghosh2020automated} & MCC & $98.33$ & - \\
\hline
$1D~ wavelet~ smoothing,~ z-score~ normalization$~\cite{alkhodari2021convolutional} & CNN-LSTM & $99.32$ & $98.30$ \\

\hline
- & \textbf{CardioXNet
(Proposed)} 	&\textbf{99.6}&	\textbf{99.4}	\\
\hline
\end{tabular}
\end{center}
\end{table*}	
\endgroup
\subsection{Comparison with the Existing Works}
A detailed comparative analysis among the proposed CardioXNet and the previous works have been presented in Table~\ref{Accuracy table}. In~\cite{khan}, manually crafted MFCC, DWT and MFCC-DWT features were extracted for feeding into the classifiers i.e., KNN, SVM and DNN. Although multiple analysis were performed adopting several feature extraction and classifier combinations, MFCC-DWT based conjoined features demonstrated the best performance, giving 97.4\%, 97.9\% and 92.1\% accuracy respectively, on KNN, SVM and DNN classifiers. Another work proposing a novel deep WaveNet architecture has achieved a training accuracy of 97\% and validation accuracy of 90\% for 5 class CVD classification~\cite{oh2020classification}.Furthermore, two recent studies on the classification of HVDs have utilized this dataset dropping the MVP class. The first work has achieved the mean accuracy of 95.13\% using wavelet synchrosqueezing transform function (WSST) for extracting the magnitude and phase features from the PCG and classified the signals with the Random Forest (RF) classifier~\cite{ghosh2019automated} while the second one contains several complex processing steps. After performing filtering and Shannon energy based cardiac cycle segmentation, local energy and local entropy features were extracted in conjunction with Chirplet transform (CT). Upon the evaluation of the features using multi-class composite classifier (MCC), the framework has obtained overall accuracy of 98.33\%~\cite{ghosh2020automated}. In another recent work CNN-LSTM model is used on PCG recordings from the same dataset with 1D wavelet smoothing and z-score normalization achieving relatively high accuracy and F1 score of 99.32\% and 98.30\% respectively. Therefore, it is clearly evident that the proposed CardioXNet architecture outperforms or performs on par with the previous works, achieving an accuracy of 99.6\%.
Individual class-wise evaluation metrics also show similar trend of superior performance in comparison with all the existing works.
\par 
The strategies adopted in~\cite{khan, ghosh2019automated, ghosh2020automated} require several computationally complex and time-consuming hand-engineered feature extraction techniques on the PCG dataset, while the CardioXNet is an end-to-end model which is capable of directly predicting CVDs from the raw PCG signal. Thus, the proposed scheme is well-performing and highly efficient with reduced complexity and latency in terms of making automated predictions. Although the WaveNet model can optimally classify the raw PCG recordings with relatively lower training parameters, it fails to perform as well as the proposed CardioXNet despite having advanced network and deep feature extraction components. Rather, it lagged by around 9.6\% accuracy in the validation phase in comparison to our proposed model. This clearly marks the lacking of robustness of the WaveNet model. In addition, the model is ambiguously represented with no clear depiction of kernels and layers; thereby, making the network irreproducible.
\par
The case of ~\cite{alkhodari2021convolutional} is different as in that work the primary focus was the classification of HVD directly from the dataset with minimal processing. However steps such as 1D wavelet smoothing and z-score normalization was performed while in our proposed work raw PCG signals were used in the model. Moreover the proposed model in  ~\cite{alkhodari2021convolutional} is vague and important details such as architecture and number of  parameters is absent. Whereas, detailed architecture of CardioXNet and its computational complexity is detailed intricately in this paper.  
\par
From this exhaustive analysis, it can be inferred that the proposed CardioXNet manifests significant improvement over all the exsting state-of-the-art works using the GitHub PCG database.

\subsection{Computational Efficiency of the Proposed Model}
In this part, a detailed bench-marking is performed on the CardioXNet,considering trainable parameters,floating point operations per second (FLOPS) required end-to-end time and size after training the parameters. The proposed lightweight CNN model has extremely low end to end time of 54.60($\pm$0.06) ms. Since the proposed paradigm can directly classify the raw PCG recordings requiring no preprocessing and  hand crafted feature extraction technique which evidently demonstrates the robustness of the model. Moreover, the proposed model has 0.67 M trainable parameters, 26 M FLOPS and a smaller memory requirement of only 7.96 MB. Therefore, it can achieve trade-off between higher accuracy than the considered networks while ensuring the significantly lower necessity of storage space and processing power. This makes the CardioXNet more suitable for real-time wearable and mobile applications~\cite{nikouei2018}.

\section{Limitations and Scopes for Future Works} \label{IV}
Although the current framework demonstrates satisfactory performance on two well-known PCG datasets representing real-life scenarios, the work can be further improved if multiple larger PCG datasets are available with diverse CVD annotations. The number of PCG data of HVD is also limited and patient number and independence is not provided within the dataset. 
\par 
Several advanced network strategies like Generative Adversarial Networks (GAN), Variational Autoencoder (VAE) can be attempted to obtain further optimized performances with more efficiency and non-latency. In future, we aim to integrate our CardioXNet in digital stethoscopes or wearable devices with a cloud server connection for performing automatic PCG classification and predicting on different CVDs using our pre-trained model in real time to assist the clinicians in their diagnostic decision.

\section{Conclusion}  \label{V}
In this work, a novel lightweight CRNN network, CardioXNet, has been proposed for automatic detection of different cardiovascular disease classes without performing any preprocessing steps on the PCG signal. The model involves both representation learning and sequence residual learning by utilizing both the CNN and bi-LSTM layers to extract time-invariant and temporal features. The framework demonstrated state-of-the-art  performance with an accuracy score of 99.6\% outperforming all the previous works on the given PCG dataset by a considerable margin. Besides showing superior results, the use of the proposed end-to-end network with significantly low number of parameters make this model well-qualified for usage in any embedded system applications. We believe that this work can contribute in the advancement of real-time, automated CVD classification from PCG recordings and can have significant real life impacts in the domain of clinical diagnostics.

\section*{Acknowledgment}
(Samiul Based Shuvo and Shams Nafisa Ali co-first authors.)

\bibliographystyle{IEEEtran}
\bibliography{ref}

\begin{IEEEbiography}[{\includegraphics[width=1in,height=1.25in,clip,keepaspectratio]{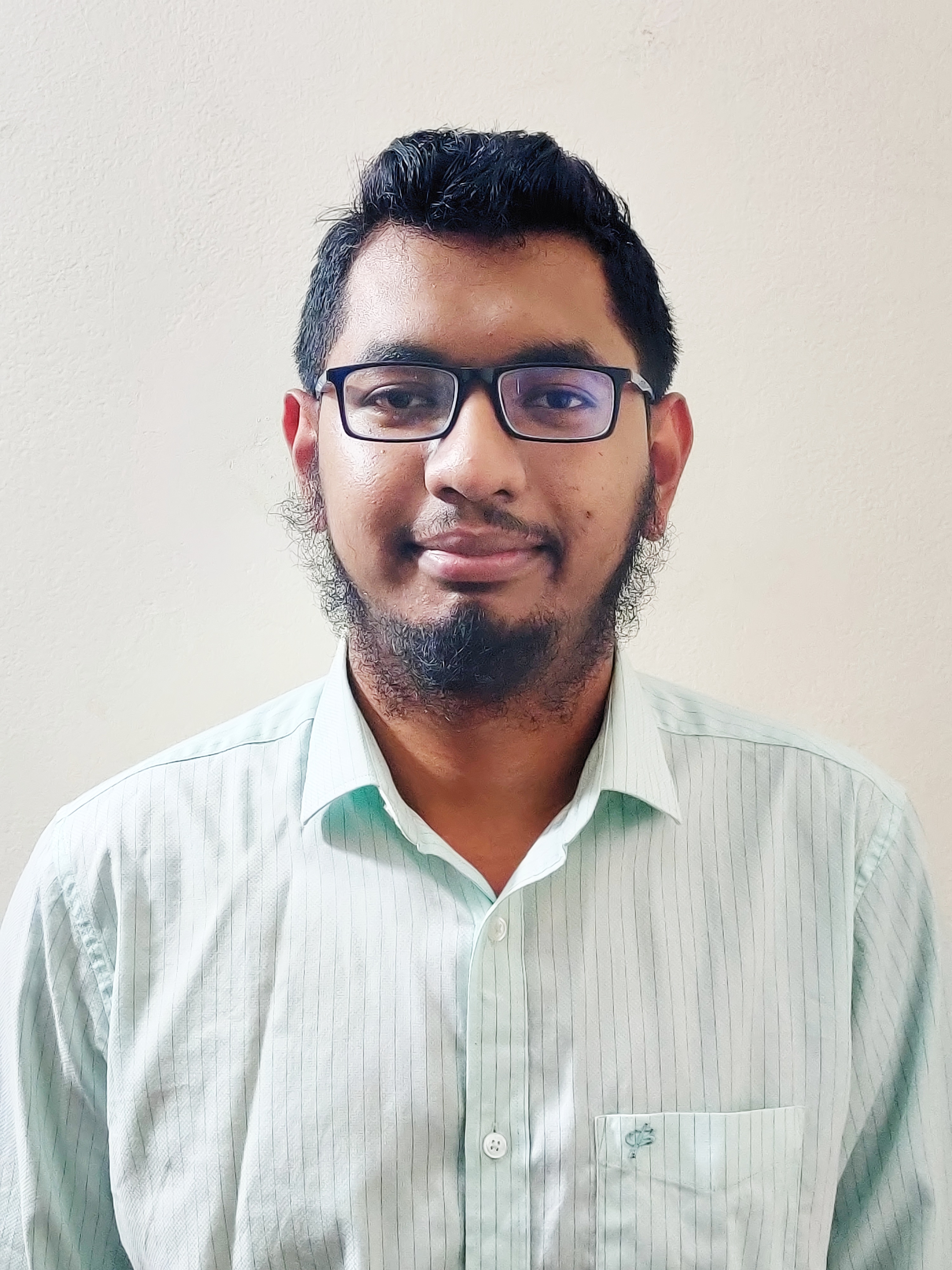}}]{Samiul Based Shuvo} is currently pursuing the B.Sc. degree in biomedical engineering at Bangladesh University of Engineering and Technology, Bangladesh. He is currently working on biomedical signal processing and deep learning-based health informatics projects. His other  research interests include biomedical simulation, biomedical Instrumentation and edge device optimization.
\end{IEEEbiography}

\begin{IEEEbiography}[{\includegraphics[width=1in,height=1.25in,clip,keepaspectratio]{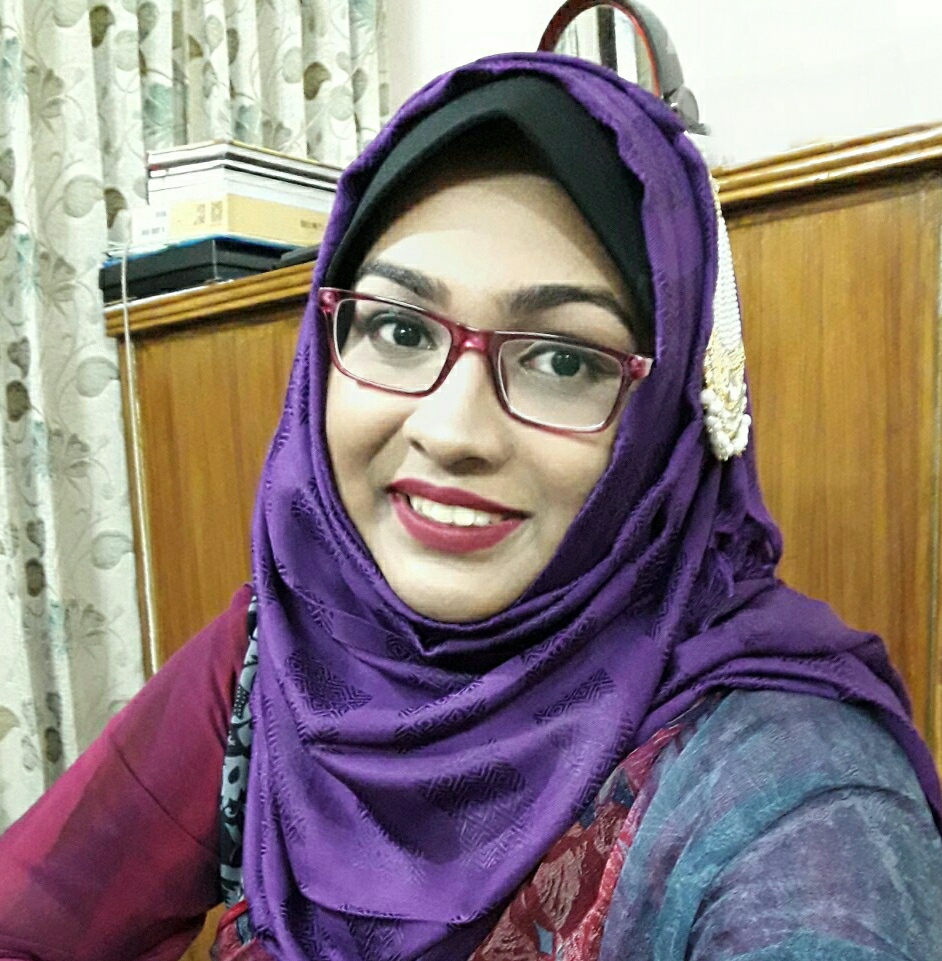}}]{Shams Nafisa Ali} (Student Member, IEEE) is currently pursuing the B.Sc. degree in biomedical engineering at Bangladesh University of Engineering and Technology, Bangladesh. Her current research interests include biomedical signal and image processing, computer aided diagnosis, deep learning, medical imaging and smart wearable systems. She is also interested to pursue the aspects of biocompatible materials through the involvement of artifical intelligence. 
\end{IEEEbiography}

\begin{IEEEbiography}[{\includegraphics[width=1in,height=1.25in,clip,keepaspectratio]{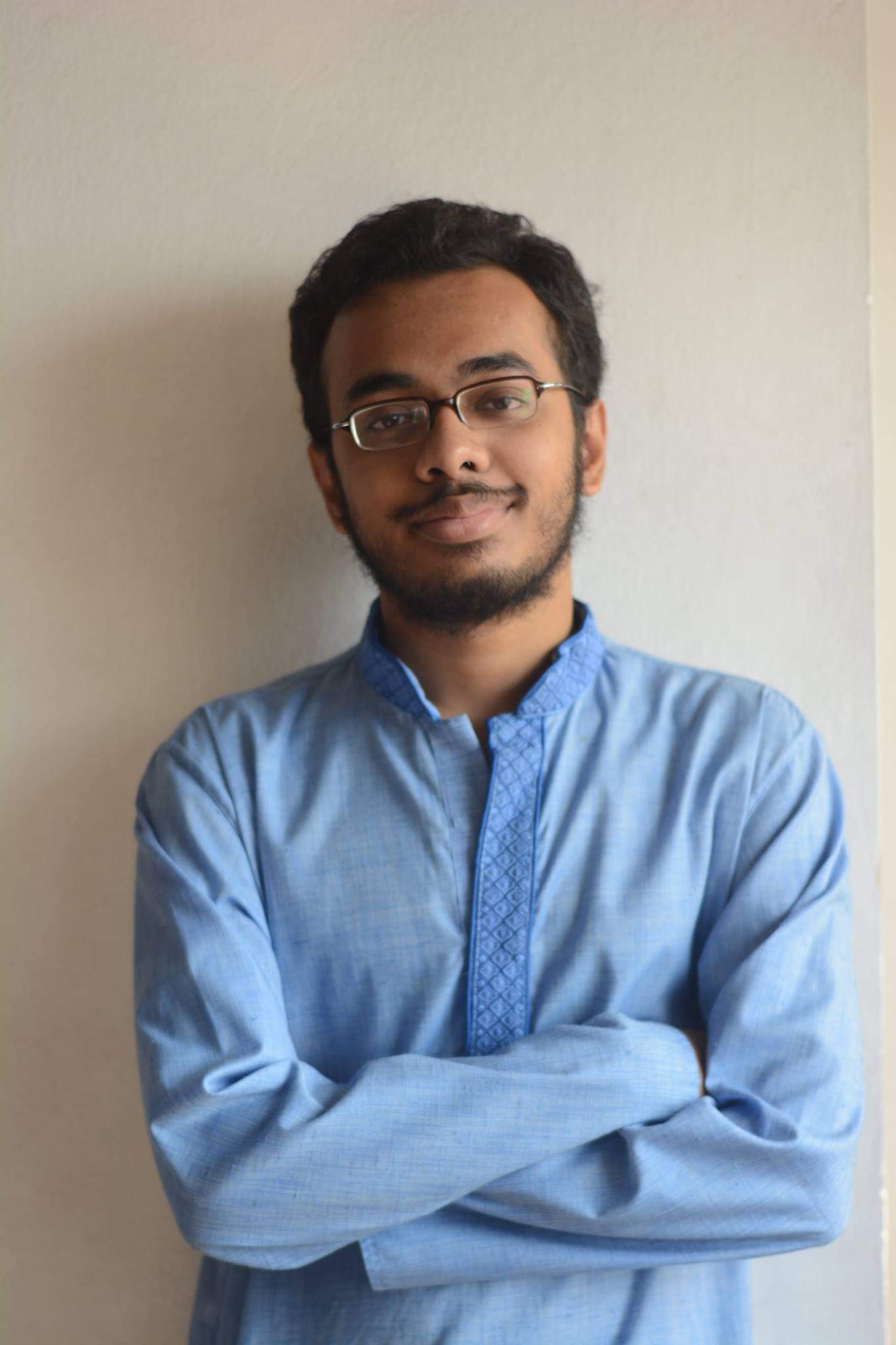}}]{Soham Irtiza Swapnil} is currently pursuing the B.Sc. degree in biomedical engineering at Bangladesh University of Engineering and Technology, Bangladesh. He is currently working on using Signal Processing and Machine Learning on biomedical signals for use in low resource settings. His other interest lies in the world of material sciences and using engineering knowledge for creating better health care solutions.
\end{IEEEbiography}

\begin{IEEEbiography}[{\includegraphics[width=1in,height=1.25in,clip,keepaspectratio]{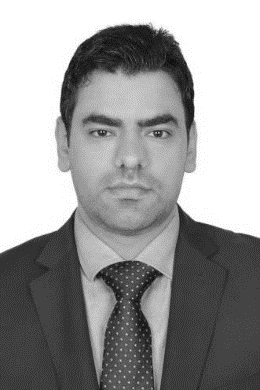}}]{MABROOK S. AL-RAKHAMI} (M’20–SM’16) received a master’s degree in information systems from King Saud University, Riyadh, Saudi Arabia, where he is currently pursuing a Ph.D. degree with the Information Systems Department, College of Computer and Information Sciences. He has worked as a lecturer and taught many courses, such as programming languages in computer and information science, King Saud University, Muzahimiyah Branch. He has authored several articles in peer-reviewed IEEE/ACM/Springer/Wiley journals and conferences. His research interests include edge intelligence, social networks, cloud computing, internet of things, big data and health informatics.
\end{IEEEbiography}
\begin{IEEEbiography}[{\includegraphics[width=1in,height=1.25in,clip,keepaspectratio]{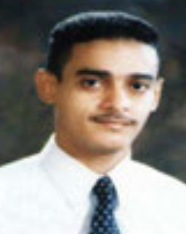}}]{ABDU GUMAEI} received the Ph.D. degree in computer science from King Saud University, in 2019. He worked as a Lecturer and taught many courses, such as programming languages at the Department of Computer Science, Taiz University, Yemen. He is currently an Assistant Professor with the College of Computer and Information Sciences, King Saud University, Riyadh, Saudi Arabia. He has authored and coauthored more than 30 journal and conference papers in well-reputed international journals. He received a patent from the United States Patent and Trademark Office (USPTO) in 2013. His research interests include software engineering, image processing, computer vision, machine learning, networks, and the Internet of Things (IoT).
\end{IEEEbiography}
\EOD
\end{document}